\documentclass[3p,times,twocolumn]{elsarticle}

\usepackage{ecrc}

\volume{00}

\firstpage{1}

\journalname{Nuclear Physics B Proceedings Supplement}

\runauth{T. Montaruli et al.}

\jid{nuphbp}

\jnltitlelogo{Nuclear Physics B Proceedings Supplement}

\usepackage{amssymb}

\usepackage[figuresright]{rotating}

\begin{document}

\begin{frontmatter}



\dochead{}

\title{Neutrino Physics and Astrophysics with IceCube}


\author{Teresa Montaruli for the IceCube Collaboration}

\address{D\'epartment de Physique Nucleaire et Corpusculaire, Universit\'e de Gen\`eve, 24 Quai Ernest Ansermet, CH-1205, teresa.montaruli@unige.ch}

\begin{abstract}
In this contribution we summarize the selected highlights of IceCube in the domain of high-energy astrophysics and particle physics.
We discuss the highest-energy neutrino detection and its interpretation after 4 years of data. The significance is such that the discovery of a non terrestrial component can be claimed but
its origin is not yet clarified.
The high energy non-atmospheric component is seen also in other analyses with smaller significance, for instance when using muon neutrino induced events coming from the Northern hemisphere.
Flavor mixing is probed along cosmic distances in an analysis using also cascade neutrino events.
The results on the search for neutrino sources is presented including the results of a joint analysis with Pierre Auger and Telescope Array which is sensitive to correlations between highest energy neutrinos and UHECRs measured by the three experiments.
Moreover, recent results on dark matter searches from the Sun are discussed. 
Finally, the results on standard neutrino oscillations are presented.
\end{abstract}


\end{frontmatter}


\section{Introduction}
\label{sec:intro}
The IceCube observatory hosts the first neutrino telescope of cubic kilometer dimensions deep in the South Pole ice between 1.45 and 2.45 km, a surface array IceTop and an in-fill array called Deep Core.
The in-ice array is running in its full configuration since the austral summer 2010-2011 and measures about 2.5 kHz of down-going atmospheric muons with an overall duty factor larger than 99\% and about 98\% of
the DOM still functioning. More than 97\% of the acquired data are
used for data analysis. 

The array is made of 5,160 Digital Optical Modules (DOMs) consisting of 10Ó photomultiplier tubes \cite{pmt} and associated
readout electronics \cite{daq} in a protective glass sphere. 
They are located along 86 strings with power and data transmission cables. 
Most of the strings are deployed at the vertices of a triangular array and horizontally spaced by about 125 m. DOMs along strings are vertically separated by 17 m. Eight of these strings  
form Deep Core, the denser in-fill array that pushes the energy threshold down to about 20 GeV of neutrino energy \cite{deepcore}. 
Additional 324 DOMs are installed in the surface array, IceTop, composed of 81
two-tank stations.The science targets of IceTop are covered in another paper of these proceedings \cite{gaisser}.

IceCube science goals are diverse: the identification of the cosmic ray sources using high-energy neutrinos as messengers, the indirect 
search for dark matter and other exotica, such as monopoles, and the measurements of neutrino oscillations and study of particle interactions in the atmosphere are some of the prominent topics, some of which are addressed below. 
The IceCube Collaboration is composed by about 300 members and 45 Institutions.

The topologies of events detected by IceCube are: tracks, characteristics of atmospheric muons which can penetrate IceCube depths or muons induced by neutrinos in the vicinity or in the instrumented ice volume;
 cascades, induced by showers produced by charged or neutral current interactions of neutrinos. The Cherenkov light is produced by ultra-relativistic charged particles at an angle of about $41^\circ$. Photons are scattered 
 in the ice with scattering length depending on depth and can propagate hundreds of meters~\cite{dima}.

\begin{figure}[htb]
          \includegraphics[width=0.45\textwidth]{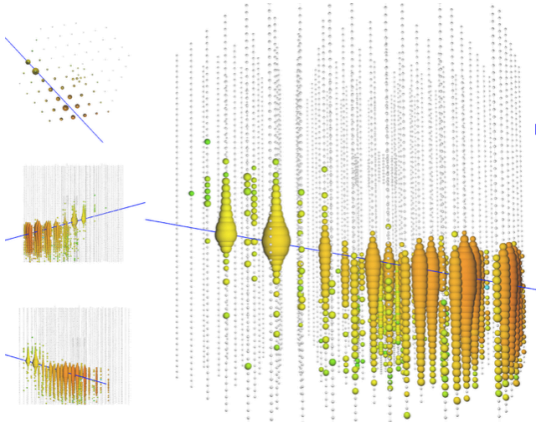}
    \caption{\label{fig:pevevent} The event of highest visible energy of $2.6 \pm 0.3$ PeV observed by IceCube. The muon releases the energy
in passing through the detector and the parent neutrino has higher energy than this. Different views of the detector are shown. The 
blobs are the DOMs and their size is proportional to the detected number of photoelectrons, while the color indicates the time. } 
\end{figure}

Once the detector has reached its final configuration, neutrino events with unprecedented energy were measured at a rate of 1 per year. 
These events are produced by neutrinos of PeV and larger energies.
Recently, the highest energy one appeared in an Astronomer's Telegram \cite{atel} and it is shown in Fig.~\ref{fig:pevevent}. 
The study of these events led to the discovery of cosmic neutrinos \cite{hese3,hese}. 

\section{The first neutrino messengers of cosmic sources}
\label{sec:cosmic}

Once the neutrino telescope has become large enough, first neutrino events from the cosmos appeared in IceCube as gigantic cascades or tracks.
They firstly showed up in a search for ultra-high energy events looking for cosmogenic neutrinos produced by ultra-high energy cosmic rays (UHECR) 
interacting with the cosmic microwave background (EHE search in Ref.~\cite{aya}). 

\subsection{IceCube searches for diffuse fluxes}
\begin{figure}[htb]
    \includegraphics[width=0.48\textwidth]{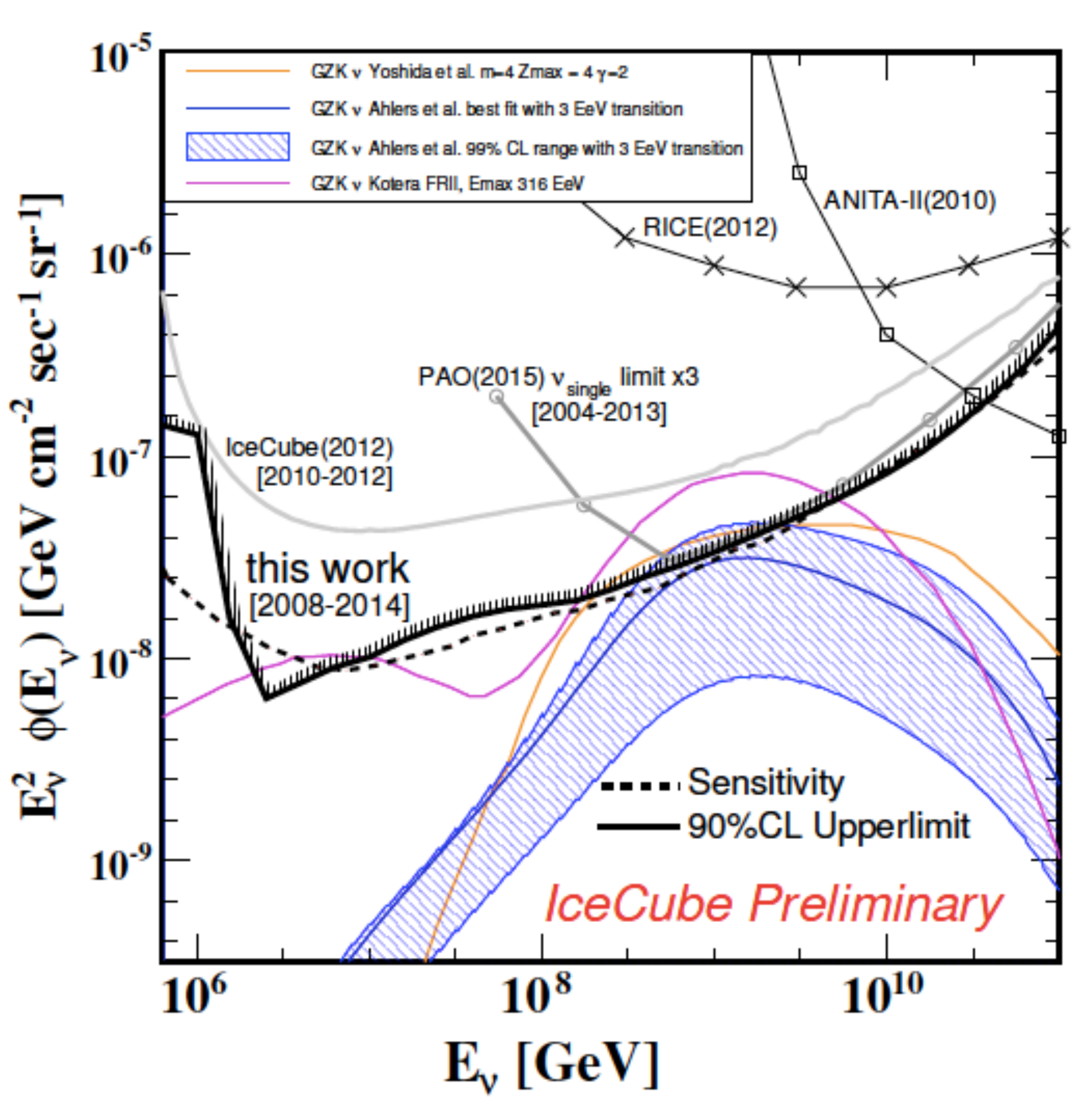}
    \caption{\label{fig:cosmogenic} 90\% CL upper limit and sensitivity for the EHE search of neutrino events for 7 years of IceCube. The equipartition of the 3 neutrino flavor fluxes at earth is assumed.  } 
\end{figure}

The IceCube search for EHE neutrino events uses the measured charge which should be larger than 30,000 photoelectrons in the detector and the angle of a line fit reconstruction of the tracks of neutrino secondaries~\cite{aya}.
One event with deposited energy of $770 \pm 200$ TeV was observed in 2050 days. This is yet incompatible with the hypothesis that these neutrinos are of
cosmogenic origin. Hence an upper limit on the fluxes of cosmogenic neutrinos produced in interactions of UHECRs with the cosmic microwave background (CMB) is set and shown in Fig.~\ref{fig:cosmogenic}.
IceCube disfavors the parameter space of UHECR sources of which the cosmological evolution is stronger than the star formation rate, where the
source candidate classes of active galactic nuclei (AGN) and gamma-ray bursts (GRB) belong under the assumption that the cosmic-ray composition is proton dominated.
This is compatible with Auger results, as discussed in \cite{aloisio}.
Nonetheless, photo-hadronic interactions are efficient only in the case of protons, while for heavy nuclei they are suppressed. Hence,
if the highest energy tail of UHECRs is composed mainly of heavy nuclei, as in models reproducing Auger data on spectrum and chemical composition, then
the flux of cosmogenic EeV neutrinos is far below the detection threshold of any running or planned detector.

An analysis was then designed to be sensitive to neutrinos of energy larger than 30 TeV by means of a veto technique that identifies High-Energy Starting Events (HESE) \cite{hese,hese3}. 
Reaching low energy, where power law fluxes are higher, and still being able to tell an atmospheric neutrino event and an astrophysical one is the challenge of cosmic sources. Energy, direction and veto information can be used for this scope.
By defining a fiducial volume through-going down-going events can be rejected and the veto efficiency can be measured on the data by using different layers of DOMs.
The remaining background, the atmospheric neutrinos, can be tagged by means of the muon brothers in the meson decays while cosmic neutrinos arrive unaccompanied \cite{schoenert}.

The HESE analysis is sensitive to neutrinos of all flavors and incident from all directions. These events with energies above 30 TeV were observed from May 2010 to May 2014 (total of 1347 days). 
The current sample consists of 52 neutrino candidates (plus two identified background events), 17 additional events compared to what published in Ref.~\cite{hese3}. Of the total 52 events, 39 are cascades and 13 track-like events.
The measured atmospheric muon background amounts to $12.6 \pm 5.1$ atmospheric muon events and the estimated 
atmospheric neutrino component, dominated by neutrinos from prompt decays of charmed mesons, is $9.0^{+ 8.0}_{-2.2}$. This has been estimated considering 
a previously set limit on atmospheric neutrinos with 59 strings~\cite{anne}. 

\begin{figure}[htb]
    \includegraphics[width=0.48\textwidth]{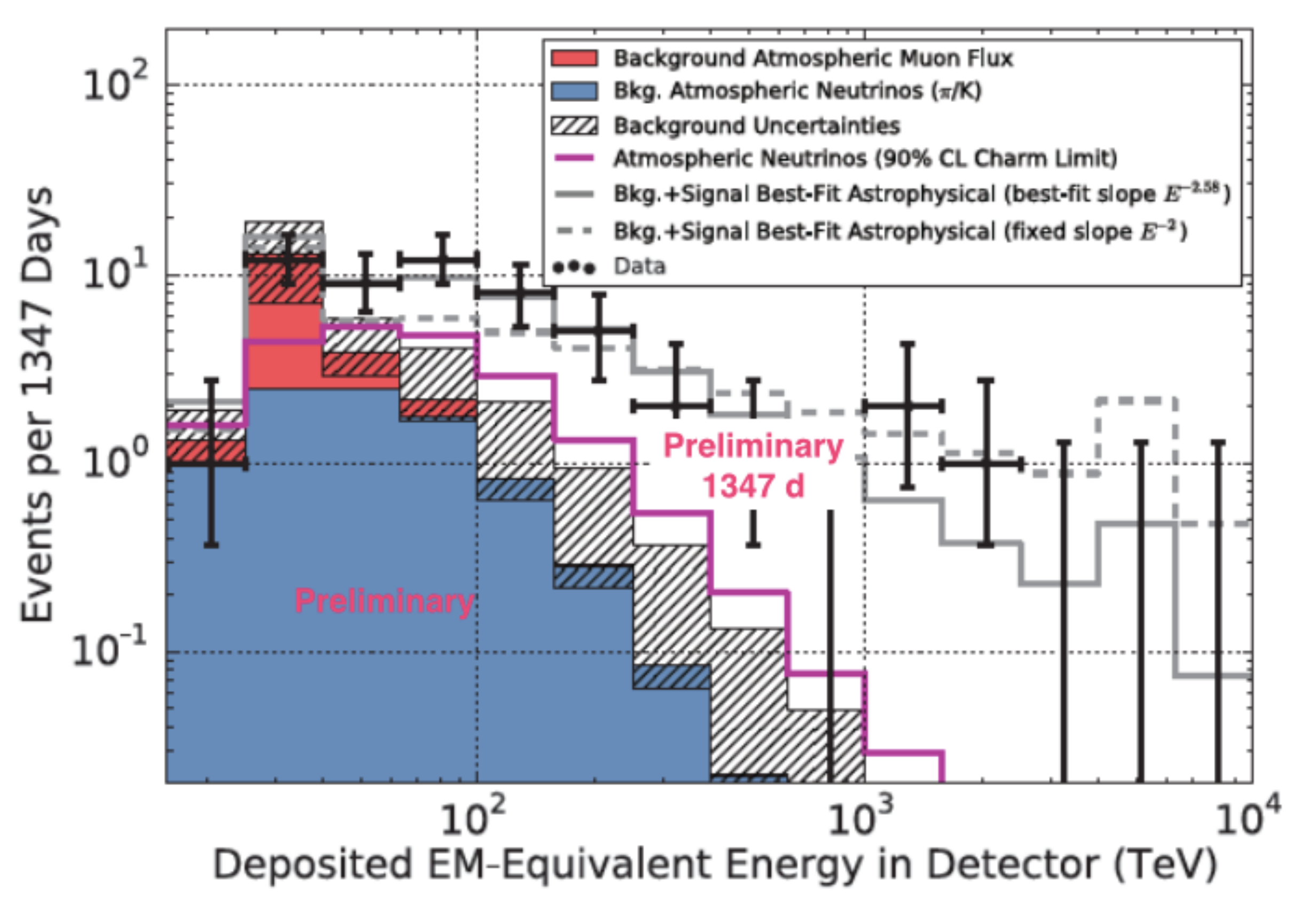}
     \caption{\label{fig:hese} The visible energy distribution in the detector of the 54 HESE in IceCube. The astrophysical component
best fit is shown with a solid line ($E^{-2.58}$ shape) while the dashed line shows an $E^{-2}$ shape as foreseen by $1^{st}$ order Fermi
acceleration mechanism. The about 21 events of atmospheric backgrounds (muons and neutrinos) are shown with filled
histograms and pink solid line, respectively. } 
\end{figure}

\begin{figure}[htb]
    \includegraphics[width=0.47\textwidth]{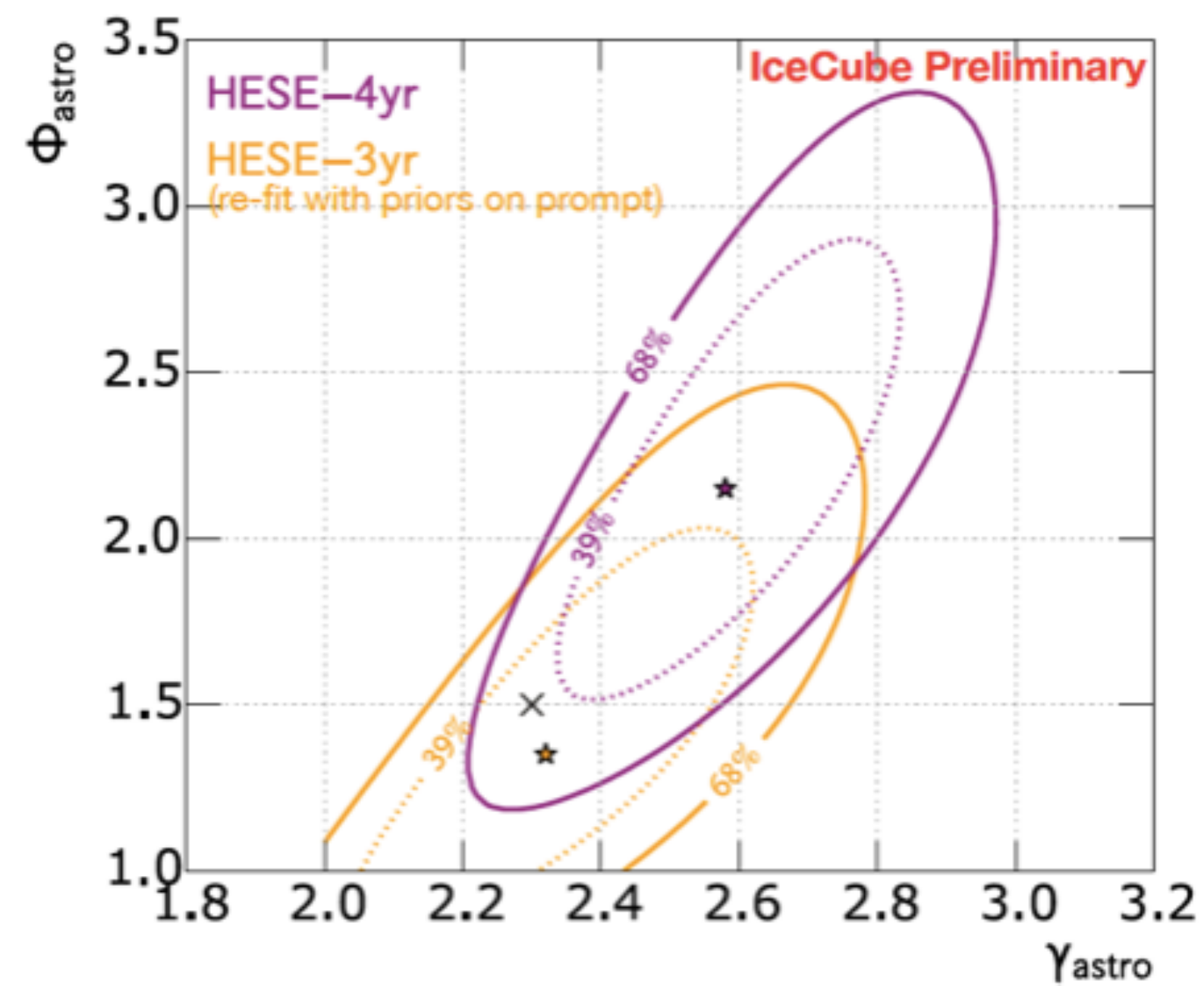}
    \caption{\label{fig:hese-spectrum} The normalization versus spectral index of the fitted astrophysical component for the 3 yr~\protect\cite{hese3} and 4 yr statistics.} 
\end{figure}

A purely atmospheric origin of these events is now excluded at a significance of 6.5$\sigma$. For an assumed unbroken canonical  $1^{st}$ order Fermi acceleration spectrum of $E^{-2}$, the resulting normalization from a likelihood fit
of all components between 60 TeV-3 PeV (atmospheric muons, atmospheric neutrinos from $K/\pi$ decays and from charmed mesons and cosmic neutrinos) is $E^2\Phi(E) = 0.84 \pm 0.3 \times 10^{-8}$ GeV cm$^{-2}$ s$^{-1}$ sr$^{-1}$.
On the other hand, when leaving free the astrophysical component spectral index, the best fit provides $-2.58 \pm 0.25$, while the published result for 3 years resulted in a bit harder cosmic neutrino spectral index:
$-2.3 \pm 0.3$. Fig.~\ref{fig:hese} shows the visible energy distribution of the HESE events, the backgrounds and the fitted astrophysical components. Fig.~\ref{fig:hese-spectrum} shows the 3 year \cite{hese3} and the updated 4 year result of the fit of the normalization versus spectral index of the astrophysical component.

Aside from cosmogenic neutrinos produced by UHECR interactions on the CMB, additionally the neutrons created in the photopion interactions decay into electron anti-
neutrinos with energies around a few PeV. Some models predict an additional neutrino flux in the lower energy region around 1 PeV which depends on higher energy infrared, optical, and ultraviolet background photons (EBL).
As pointed out in Ref.Ê\cite{aloisio}, the level of the inferred astrophysical flux by IceCube is compatible with the 
flux of cosmogenic neutrinos in the PeV energy region which is produced by the UHECRs in the low energy part of the observed spectrum (around $E \sim 10^{18.5}$ eV), for which
light composition is well established. This fact enables a constrain on cosmological evolution of sources, disfavoring models with a too
strong evolution: $S(z) \propto (1 + z)^m$ with $m \gtrsim 3.5$.

\begin{figure}[htb]
    \includegraphics[width=0.48\textwidth]{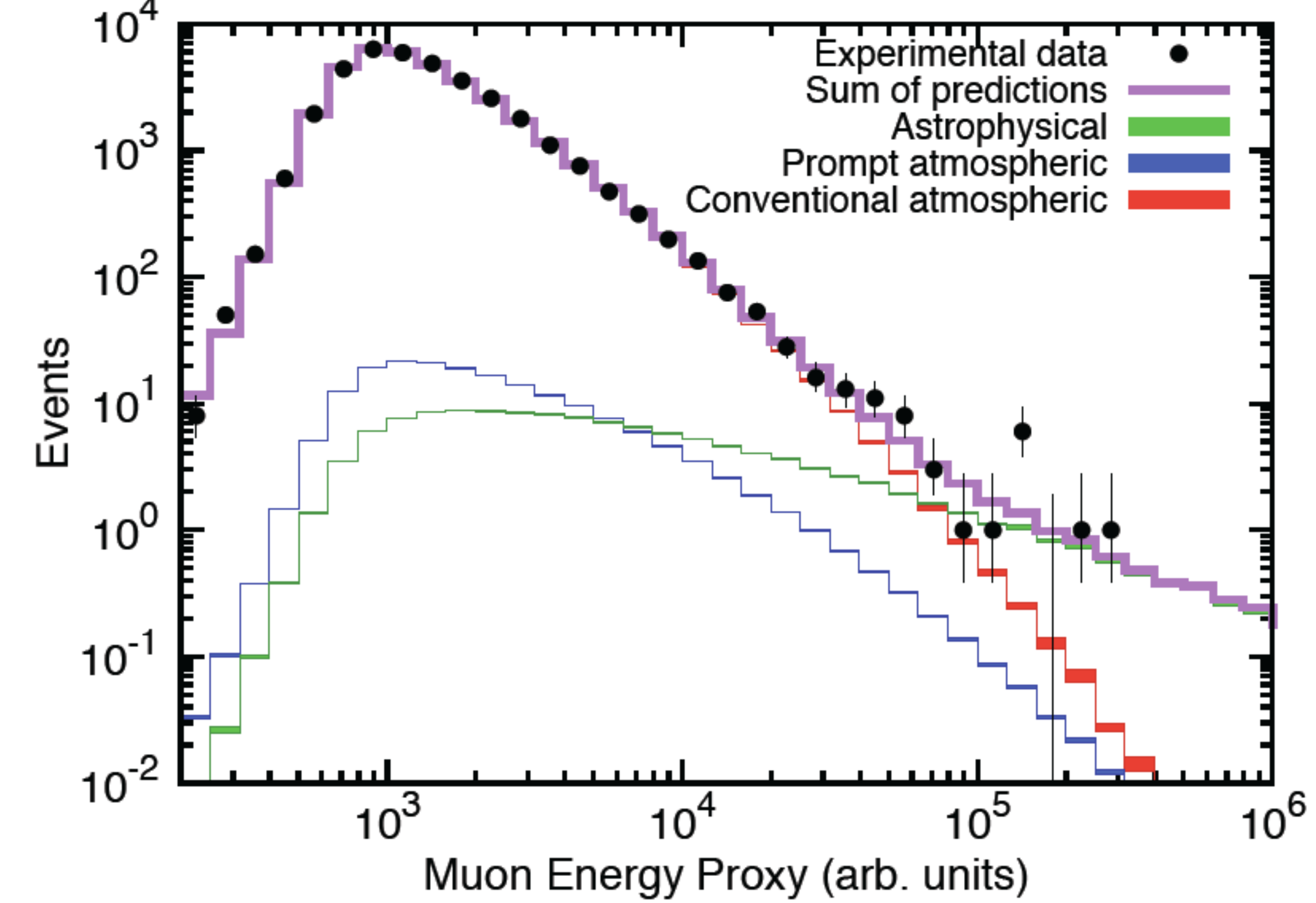}
    \caption{\label{fig:diffusenumu}  The distribution of the reconstructed muon energy proxy for up-going muon neutrino events selected in Ref.~\protect\cite{weaver}, compared to the expected distributions
for the backgrounds and an $E^{-2}$ astrophysical neutrino spectrum (green line). Only statistical errors are shown. The energy proxy does not
have a linear relationship to actual muon energy, but values of $3\times 10^3$ are roughly equivalent to the same quantity in GeV.
Larger proxy values increasingly tend to underestimate muon energies, while smaller values tend to overestimate.}
 \end{figure}

Another search for diffuse fluxes looked for high-energy up-going muon tracks produced by muon neutrinos interacting outside of
IceCube and producing up-going through-going muons~\cite{weaver}. In this case neutrinos are filtered by the earth itself but atmospheric neutrinos can only be distinguished from cosmic neutrinos
on a statistical basis, looking for deviations of the measured flux from the atmospheric muon neutrino spectrum. This analysis  produced $3.7\sigma$ evidence for astrophysical neutrinos in 2 yrs of data. The excess
of through-going muons at high energy is shown in Fig.~\ref{fig:diffusenumu}.

\begin{figure}[htb]
    \includegraphics[width=0.48\textwidth]{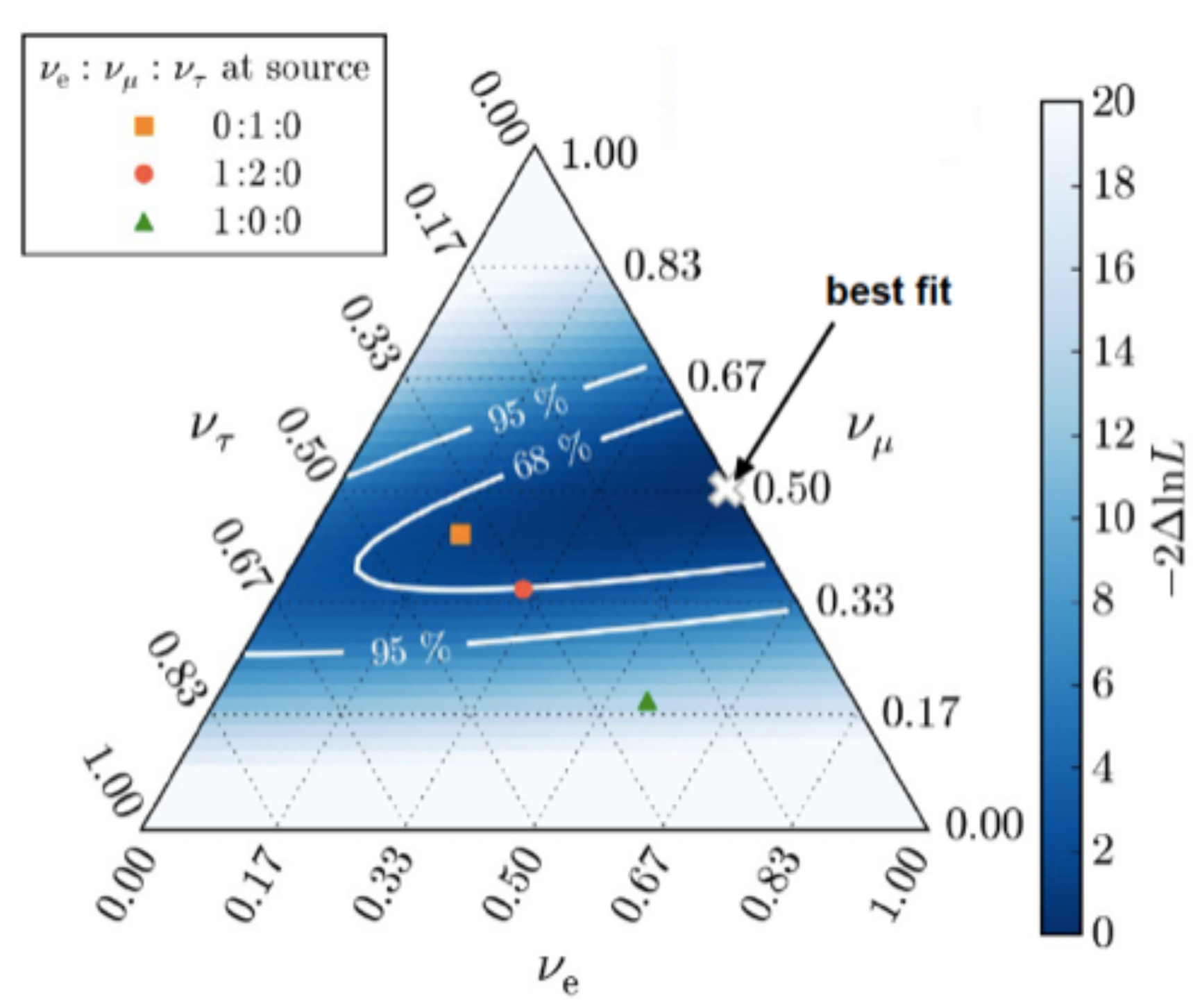}
    \caption{\label{fig:flavor} Profile likelihood scan of the flavor composition at Earth. Each point in the triangle corresponds to a ratio $\nu_e : \nu_{\mu} : \nu_{\tau}$ measured on Earth, the individual contributions
are read on the 3 sides of the triangle. The best-fit composition is marked with X, the 68\% and 95\% confidence
regions are in different tones of blue. The best-fit composition obtained in an earlier IceCube analysis of
the flavor composition \protect\cite{flavor} is marked with a +. The ratios corresponding to three flavor composition scenarios at the sources of the neutrinos are marked by:
the square (0 : 1 : 0) for the scenario where muons do not decay due to their high energies; the circle (1 : 2 : 0) for the scenario in which in the source mainly pions decay, and triangle (1 : 0 : 0) for the scenario of a source where only neutrons decay. Figure from Ref.~\protect\cite{lars}. }
 \end{figure}

Further exploration of the cosmic component was performed with a fit with 3 variables (energy, zenith angle and topology) of 8 samples from searches performed between 2008 and 2014 in the TeV-PeV energy range.
This resulted in two best fits of an unbroken and a broken astrophysical component where this last is preferred to the first one by $1.2\sigma$.\cite{lars}
An updated result on the flavor composition has also been presented compared to the initial result in Ref.~\cite{flavor} and the result is shown in Fig.~\ref{fig:flavor}. So far, with the given limited statistics, 
the observed astrophysical flux is consistent with an isotropic flux of
equal amounts of all neutrino flavors. 

\subsection{Searches for astrophysical sources of neutrinos}

\begin{figure}[htb]
    \includegraphics[width=0.48\textwidth]{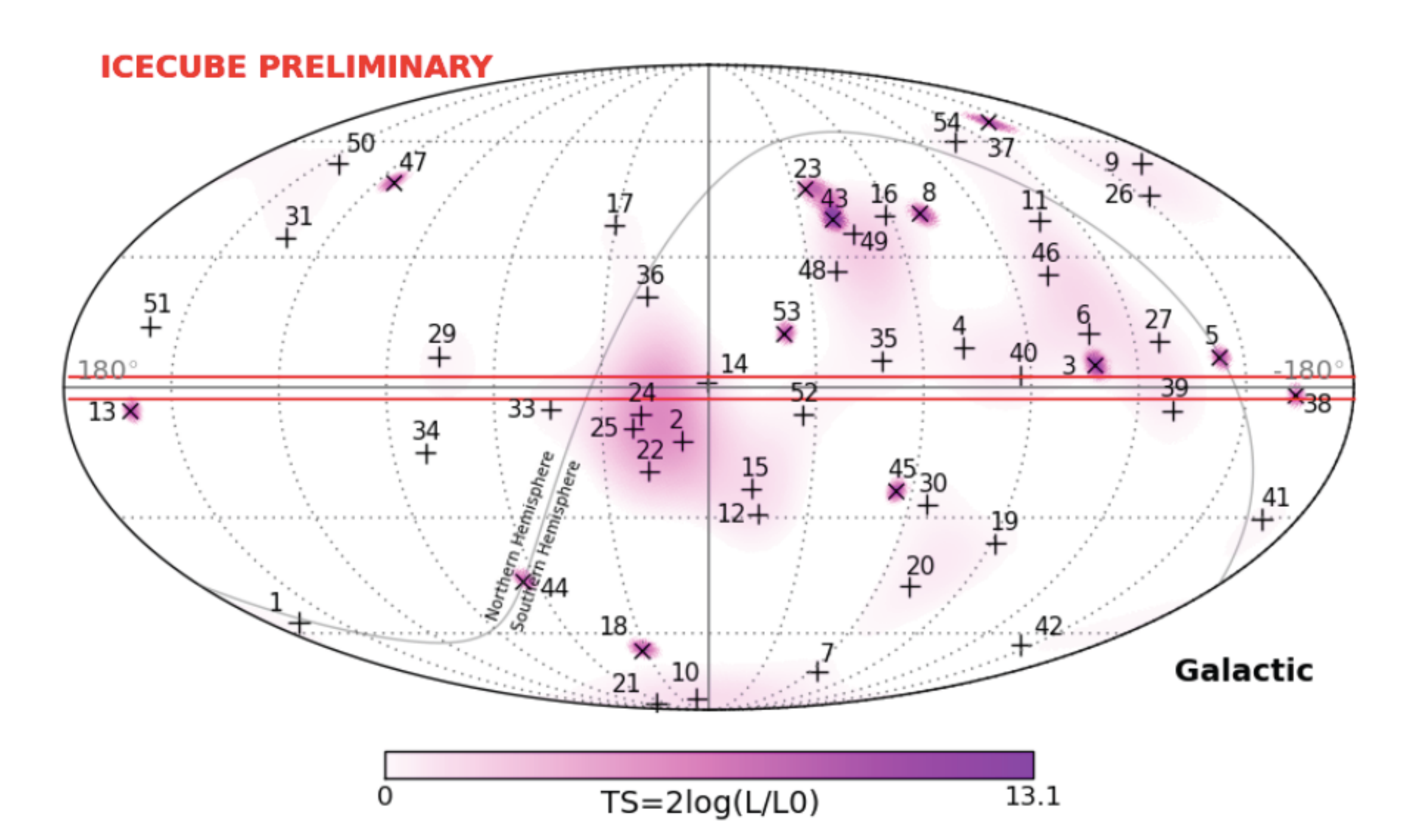}
    \caption{\label{fig:skymap} Arrival directions of the 52 high energy contained neutrino candidate events in galactic
coordinates. Crosses indicate cascade-like events while XÕs show tracks. The color scale indicates
the significance of the event excess at each point, which depends on the reconstruction uncertainty of
each event. Events 28 and 32 are identified as background and are omitted.} 
\end{figure}

The clustering of HESE events has been tested and did not yield significant evidence with post trial p-values of 44\% and 58\% for the shower-only and all-events, respectively. 
A galactic plane clustering test using a fixed width of $2.5^\circ$ around the plane and using a variable-width scan led to post trial p-values of 7\% and 2.5\%, respectively. 
A galactic coordinate sky map of the HESE events is shown in Fig.~\ref{fig:skymap}. For cascades the directional precision is poor (order of $10^\circ-20^\circ$) while for tracks 
the angular resolution is typically better than $1^\circ$.

The possibility that cosmic rays and neutrinos come from the same sources was tested.
In such a case, the angular separation between their arrival directions would be related to the magnetic deflections suffered by
ultra-high energy cosmic rays (UHECR) during their propagation, convoluted with the experimental angular resolution. This angular separation is unknown a priori due to the poor knowledge of the intervening extragalactic and galactic magnetic fields 
and the uncertain value of the composition of UHECRs.

UHECR of energies above 50 EeV are detected by two huge extensive air showers: the Pierre Auger Observatory in Argentina and the Telescope Array in Utah. The Pierre Auger surface detector (SD) is composed of 1660 water-Cherenkov stations spread
over an area of about 3'000 km$^2$ and the fluorescence detector (FD) comprises 27 telescopes at four sites detecting the fluorescence light emitted by nitrogen molecules excited by the
particles from the air showers. The Telescope Array (TA) consists of 507 plastic scintillator detectors, each of 3 m$^2$ in area, located on a 1.2 km
square grid and covering an area of approximately 700 km$^2$ and 38 fluorescence telescopes arranged
in 3 stations. A total number of 231 Pierre Auger events and  87 from telescope array were used in 
three searches for correlations between neutrino candidates detected by IceCube and the described UHECRs.

\begin{figure}[htb]
    \includegraphics[width=0.48\textwidth]{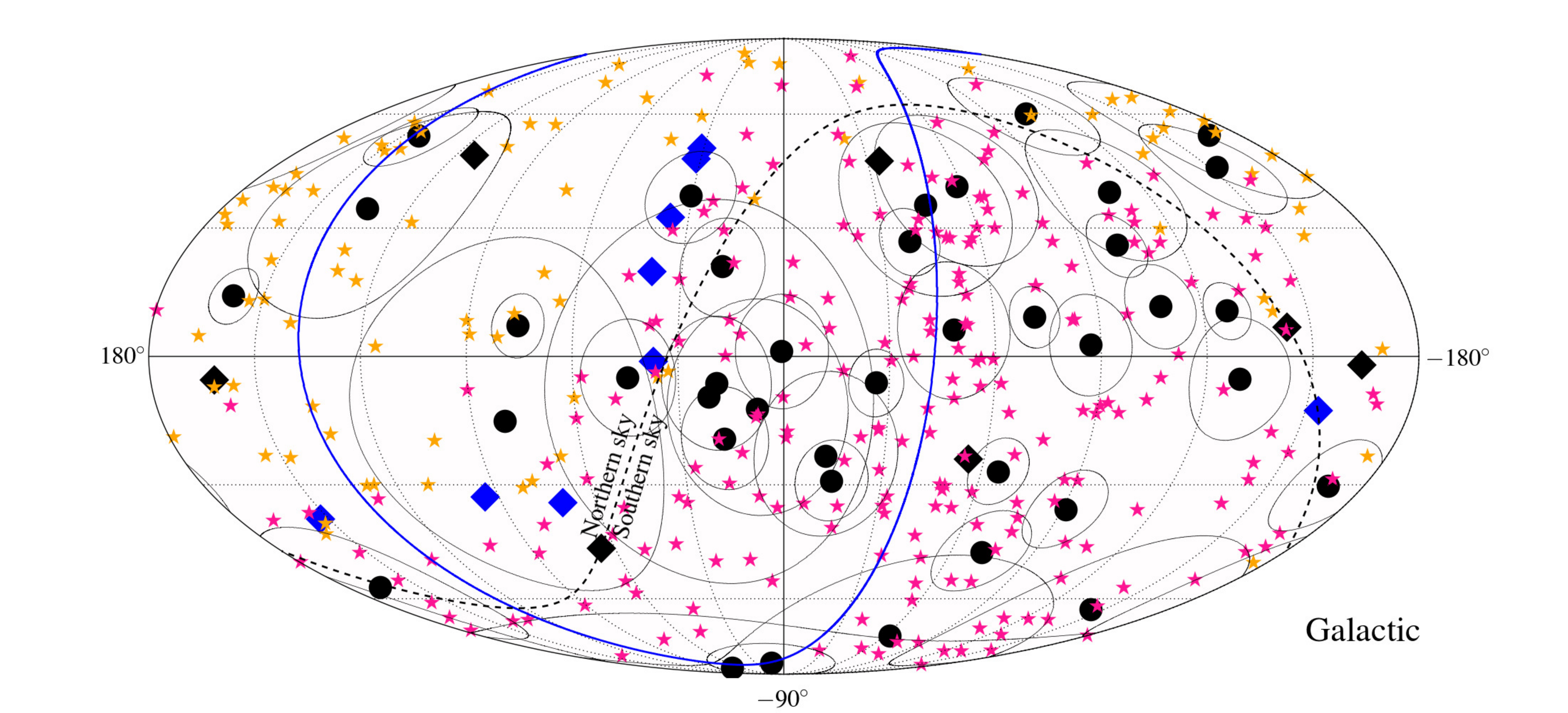}
    \caption{\label{fig:skymap_uhecr} Maps in galactic coordinates showing the arrival directions of the
IceCube cascades (circles) and tracks (diamonds), as well as those of the UHECRs detected
by the Pierre Auger Observatory (magenta stars) and TA (orange stars). The
circles around the showers indicate angular errors. The black diamonds are the HESE tracks
while the blue diamonds stand for the tracks from the through-going muon sample. The blue
curve indicates the Super-Galactic plane.} 
\end{figure}

Two of the searches, a cross-correlation analysis and a stacking likelihood one, used a sub-sample of HESE cascades (7 of the muon neutrino events which, given their topology and energy, are presumably of non-terrestrial origin, and 9 highest-energy events selected in the diffuse up-going muon neutrino search described above). Also the 39 HESE cascades were used, but given the very different angular resolution, separate p-values were derived between cascades and tracks. 
All events are shown in galactic coordinates in Fig.~\ref{fig:skymap_uhecr}. The third search uses a background contaminated sample with sub-degree angular resolution used for point source searches.

\begin{figure}[htb]
    \includegraphics[width=0.48\textwidth]{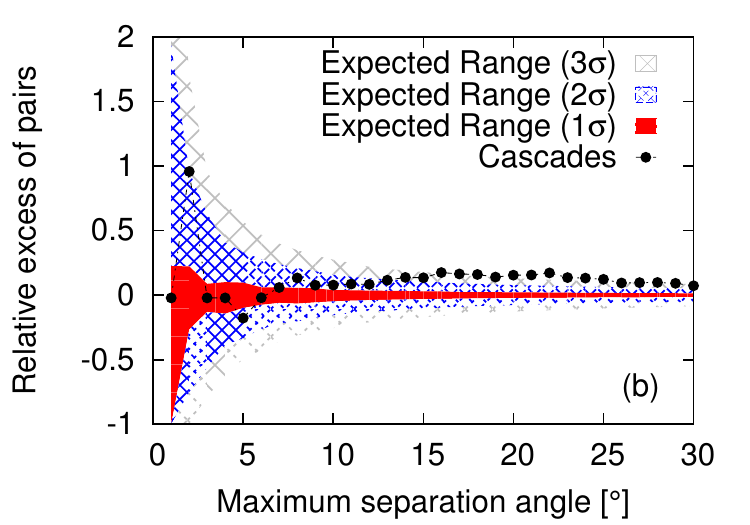}
    \caption{\label{fig:corr1} Relative excess of pairs, $[n_{p} (\alpha)/n_{p}^{iso}(\alpha)]-1$, vs the maximum angular
separation between the neutrino and UHECR pairs, for the analysis done with the cascade
events. The $1\sigma$, $2\sigma$ and $3\sigma$ fluctuations expected from an
isotropic distribution of arrival directions of cosmic rays are shown in red, blue and grey, respectively.} 
\end{figure}

The cross-correlation method consists of computing the number $n_p$ of UHECR-neutrino pairs versus their angular separation $\alpha$, and comparing it to the expectation from
an isotropic distribution of UHECR arrival directions. An angular scan was performed between $1^\circ$ and $30^\circ$ not to rely on any assumption about the exact value of the strength of the
magnetic deflections. Results are shown in Fig.~\ref{fig:corr1} for cascades.
While tracks did not indicate any deviation for an isotropic distribution, the maximum departure from the isotropic UHECR distribution for cascades (as visible in Fig.~\ref{fig:corr1}) happened at 
an angular distance of $22^\circ$ with a post-trial p-value of $5.0 \times 10^{-4}$. From the map in Fig.~\ref{fig:skymap_uhecr} it can be inferred that the excess comes mostly from the TA 'hot spot' \cite{TA}
and the direction of the powerful AGN Cen A.

Another search uses a stacking likelihood method to reduce the trial factor of the cross-correlation search at the expenses of assuming fixed deflection parameter $D = 3^\circ$ and $6^\circ$ at 100 EeV and a deflection decreasing linearly with the inverse of the UHECR energy. 
In this case the common UHECR-neutrino sources are sampled from each neutrino event likelihood map. Results of this search are compatible with the cross-correlation.

The final search using the lower-energy point source sample did not find any significance indicating a possible correlation.
Results are submitted for publication by the three Collaborations \cite{icecube-uhecr}.

\begin{figure}[htb]
    \includegraphics[width=0.48\textwidth]{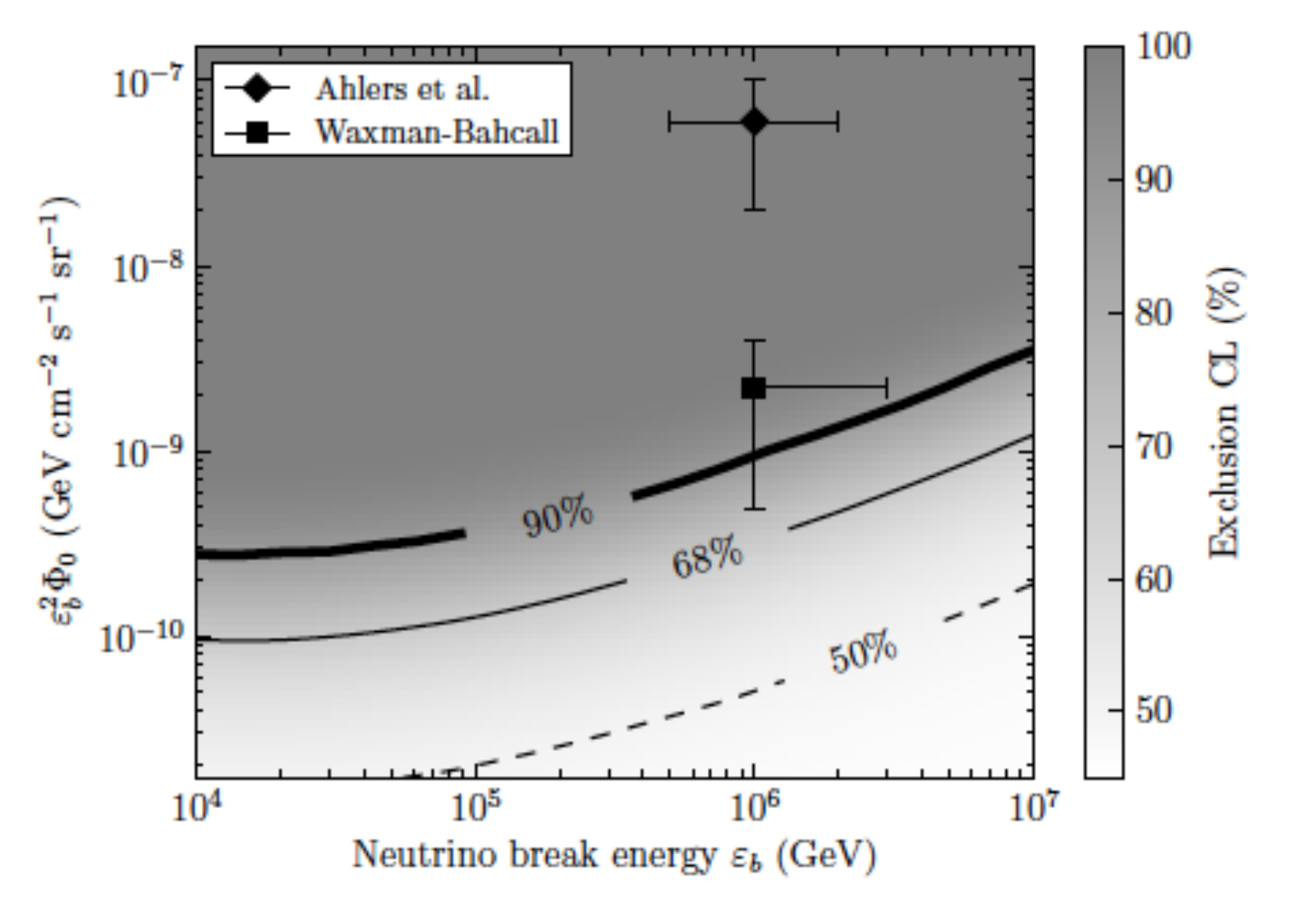}
    \caption{\label{fig:grb} Constraint on generic doubly broken power law neutrino flux models from GRBs vs first break energy $\epsilon_b$ and normalization $\Phi_0$ of the GRB flux \protect\cite{grb}. The model in Ref.~\protect\cite{ahlers} assumes that
only neutrons escape from the GRB fireball to contribute to the UHECR flux and the model in Ref.~\protect\cite{WB} allows all protons to escape the fireball.} 
\end{figure}

Many searches for time-dependent emissions of cosmic neutrinos have been performed in IceCube including target of opportunity programs. 
A search for prompt neutrino emission from gamma-ray bursts using 4 years of IceCube data has set a strong upper limit because the small
spatial and temporal search windows dramatically reduce the atmospheric neutrino background~\cite{grb}. IceCube limits increasingly constrain GRBs as dominant sources of UHECRs (see Fig.~\ref{fig:grb}).

\section{Measurement of atmospheric fluxes}

\begin{figure}[htb]
    \includegraphics[width=0.45\textwidth]{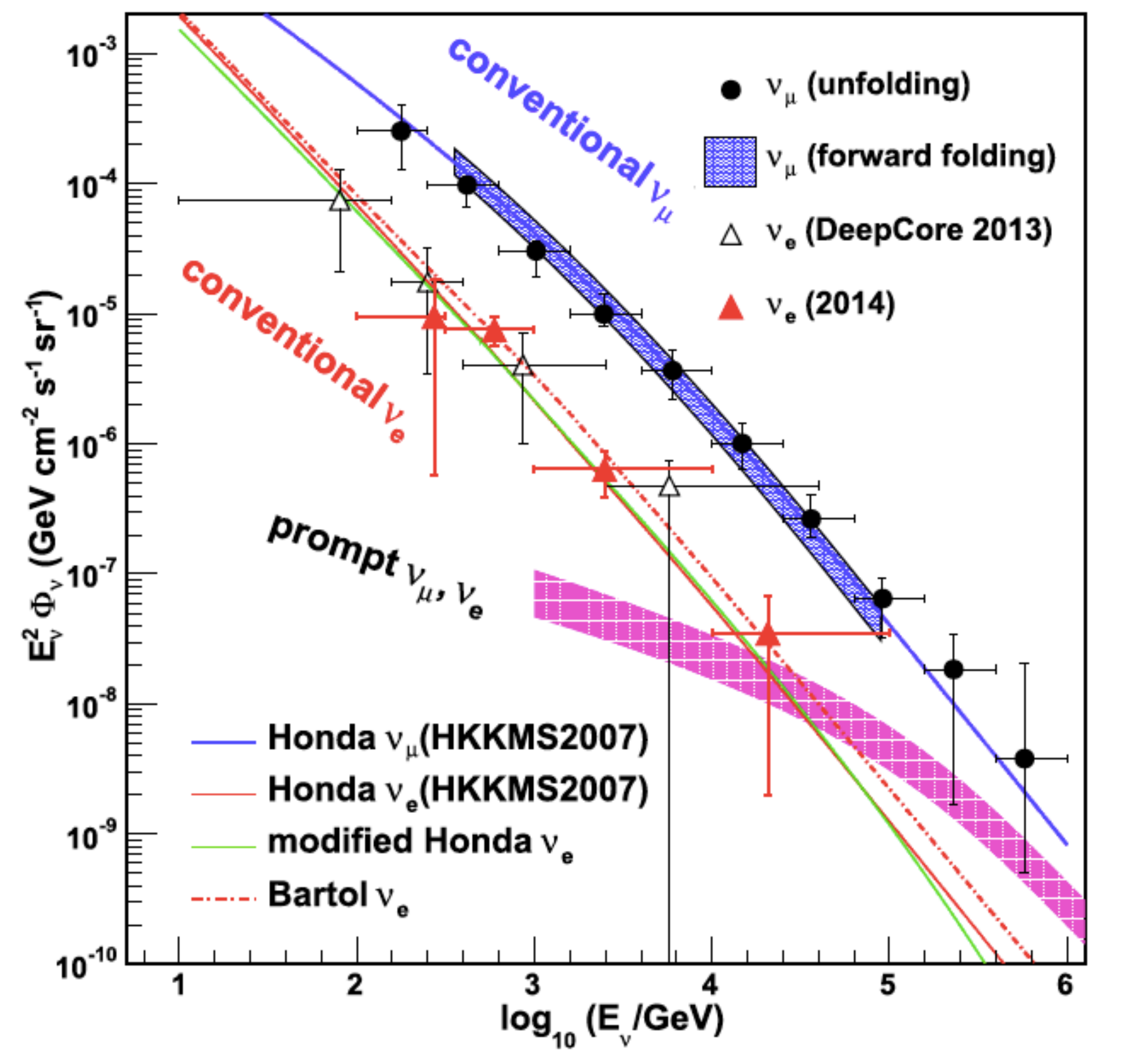}
    \caption{\label{fig:atmonu} The atmospheric $\nu_e$ flux (filled
triangles) and $\nu_{\mu}$ flux (filled circles). The open triangles show the $\nu_e$ measurement with the IceCube-DeepCore
dataset. The magenta band shows the modified ERS prediction. All references in \protect\cite{atmo_nue}.} 
\end{figure}

Having observed a significant excess over atmospheric components, it is extremely important for IceCube to pin down the nature of the excess, to identify the sources and to produce a careful measurement of the atmospheric component.
This measurement includes the discovery of the prompt neutrino and muon components from charmed meson decays and other heavy particles in order to better identify the shape of the cosmic spectrum and also to 
contribute a better understanding of high energy particle interactions. 
The measurement of the atmospheric electron neutrinos at energies $> 100$ GeV was published recently in Ref.~\cite{atmo_nue} from which the Fig.~\ref{fig:atmonu}. In this figure also the muon neutrino flux is shown. 
While the prompt component is not yet showing up, it is noticeable that improved statistics at high energy, particularly for the $\nu_e$
component which is less affected by muon background, may lead to the discovery of prompt neutrinos.

It is also extremely important to use the huge muon statistics to understand high-energy hadronic interactions and primary composition at energies above the TeV. Moreover prompt muon fluxes are enhanced with respect to neutrino ones by the electromagnetic decays of unflavored vector mesons \cite{anatoly}.
A first muon flux measurement has been submitted for publication \cite{muon} where indications of an excess of muon events which is compatible with highest pQCD predictions of a prompt component is discussed.
Such excess should be confirmed with higher statistics in the neutrino samples.

\section{The indirect search for dark matter}
\label{sec:dm}

IceCube performs multiple indirect dark matter searches from celestial regions and bodies. In the center of the Sun, WIMPs can be gravitationally trapped and pair annihilate into standard model
particles. The MSSM neutralino at non relativistic velocities annihilates preferentially
into heavy fermion-antifermion pairs such as top, bottom and charm quarks and tau leptons, as
well as heavy gauge boson pairs like $W^+W^-$ and $ZZ$. When WIMPs annihilate inside the Sun, neutrinos are the only stable Standard Model particles that can leave the
Sun without being completely absorbed. All flavors of neutrinos are produced and mix due to oscillations in the Sun and in vacuum.
The annihilation channels that produce high energy neutrinos are b, c  and t 
quarks as well as $\tau$-leptons and gauge bosons.
Scenarios in which WIMPs annihilates 100\% into $b \bar{b}$ (soft), in which the neutrino emission peaks at energies much below the WIMP mass, or 100\% into $W^+W^-$ and $\tau^+ \tau^-$ channels (hard), in which the neutrino
emission is peaked at energies close to the WIMP mass, are considered to derive upper limits. 

\begin{figure}[htb]
    \includegraphics[width=0.47\textwidth]{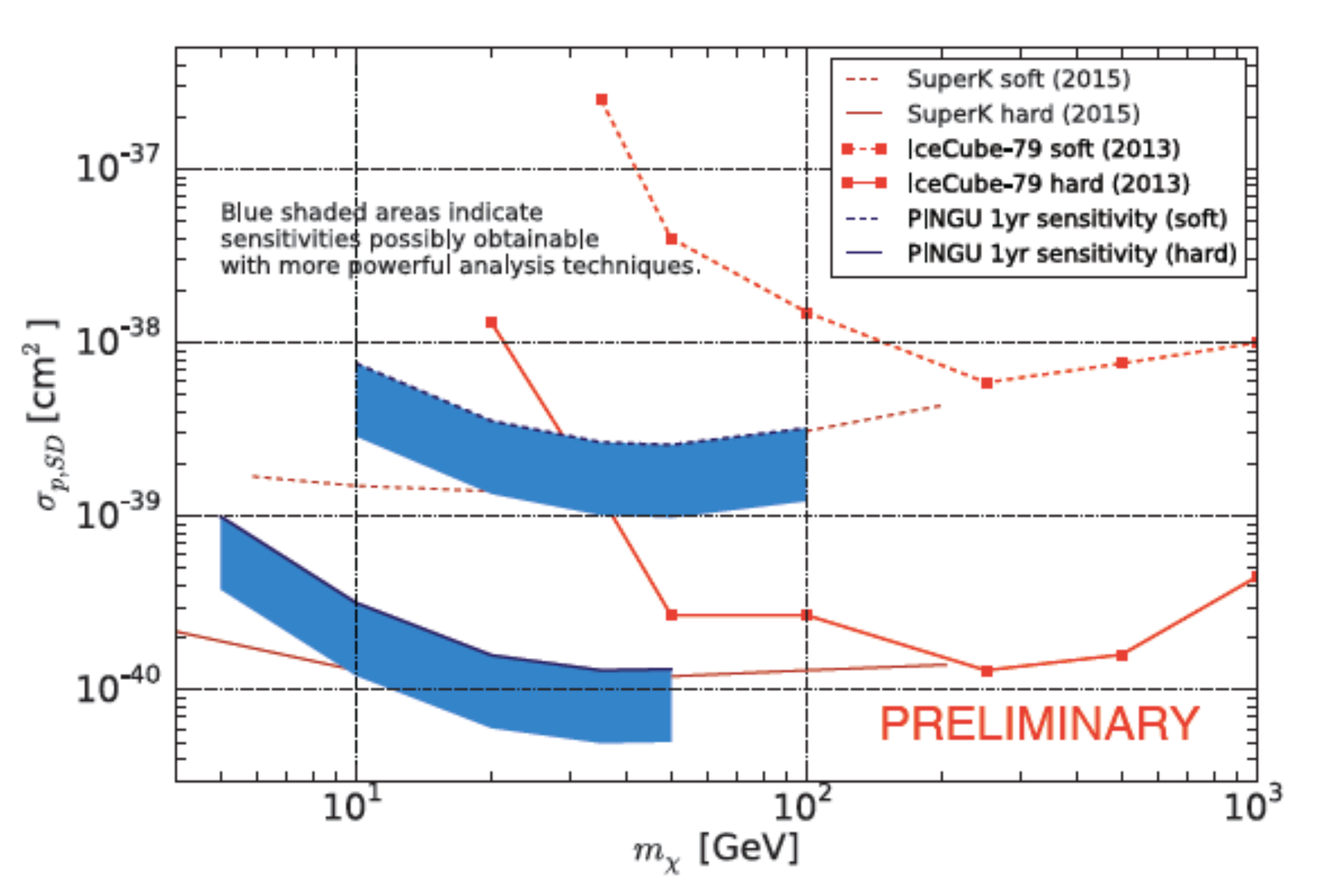}
    \caption{\label{fig:PINGU_dm} The dark matter exclusion
limit (90\% C.L.) for IceCube in its 79 configuration \protect\cite{WIMP79}, Super-Kamiokande and the projected results for PINGU \protect\cite{gen2} for 1 year of data taking.} 
\end{figure}

WIMPs produce relatively low energy events in IceCube, hence it is important the contribution of DeepCore and of the future larger in-fill array PINGU \cite{PINGU} (see Fig.~\ref{fig:PINGU_dm}). In-fill arrays allow to extend the search to summertime, when the Sun is above the horizon and
neutrinos can be tagged by recognizing their vertex in a fiducial region. Hence current IceCube searches for dark matter in the Sun extend to the full year, while in the past only upgoing event, where neutrinos are filtered by the earth where used.

\begin{figure}[htb]
    \includegraphics[width=0.46\textwidth]{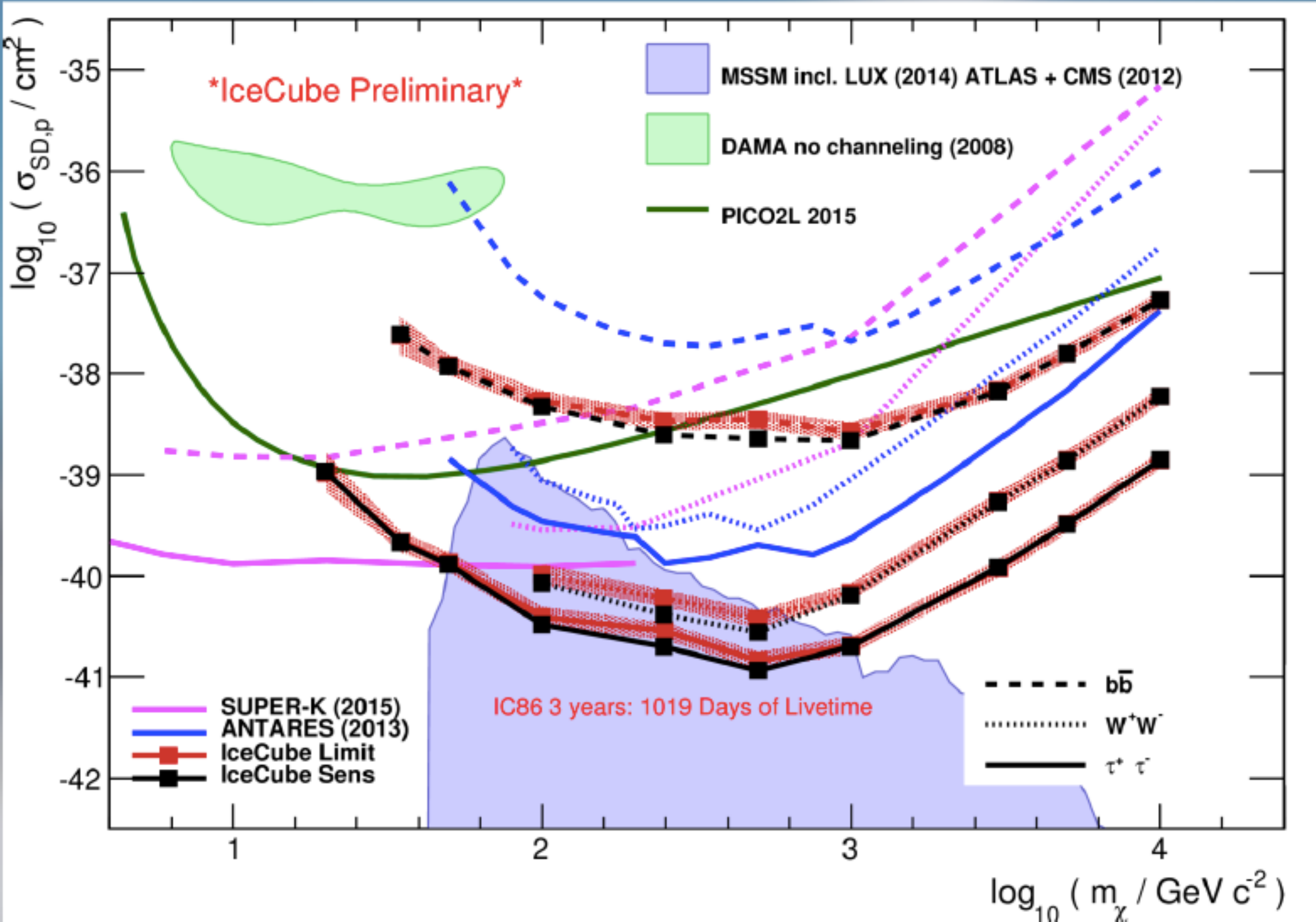}
    \caption{\label{fig:wimps_sd} 90\% CL limits on the spin dependent neutralino-proton cross section, compared to results from other neutrino detectors and direct detection experiments.} 
\end{figure} 

\begin{figure}[htb]
    \includegraphics[width=0.47\textwidth]{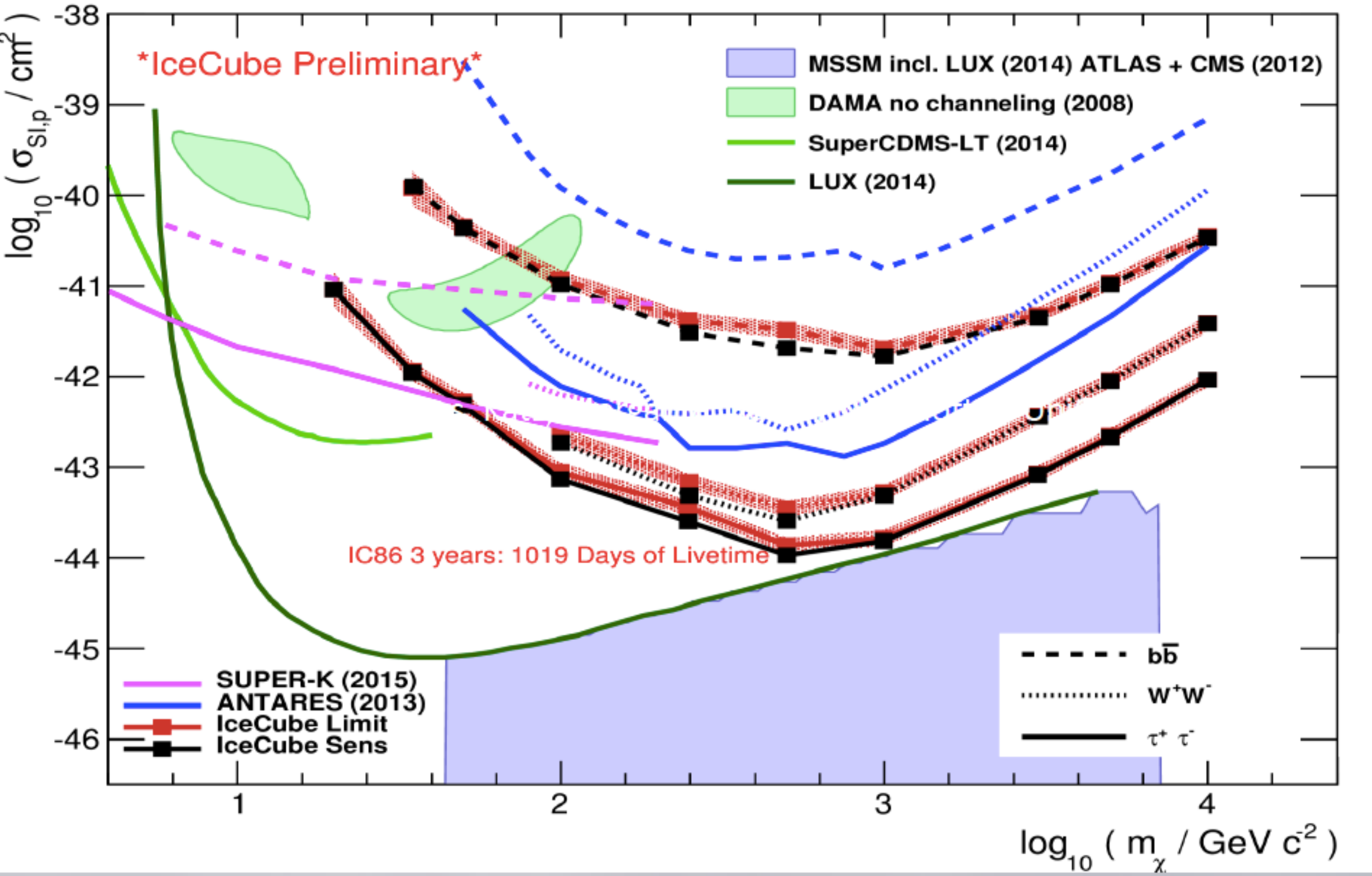}
    \caption{\label{fig:wimps_si} Same as Fig.~\protect\ref{fig:wimps_sd} for spin independent.} 
\end{figure}

Recently 3 years of full IceCube have superseeded previous limits \cite{WIMP79}.
Limits on the spin dependent and spin independent wimp-proton cross section are shown in Fig.~\ref{fig:wimps_sd} and Fig.~\ref{fig:wimps_si}. For spin dependent models, IceCube has unprecedented 
sensitivity and it is also quite sensitive at large WIMP masses for the spin independent case, complementing direct and collider searches.

\section{Neutrino oscillation results from IceCube}

\begin{figure}[htb]
    \includegraphics[width=0.47\textwidth]{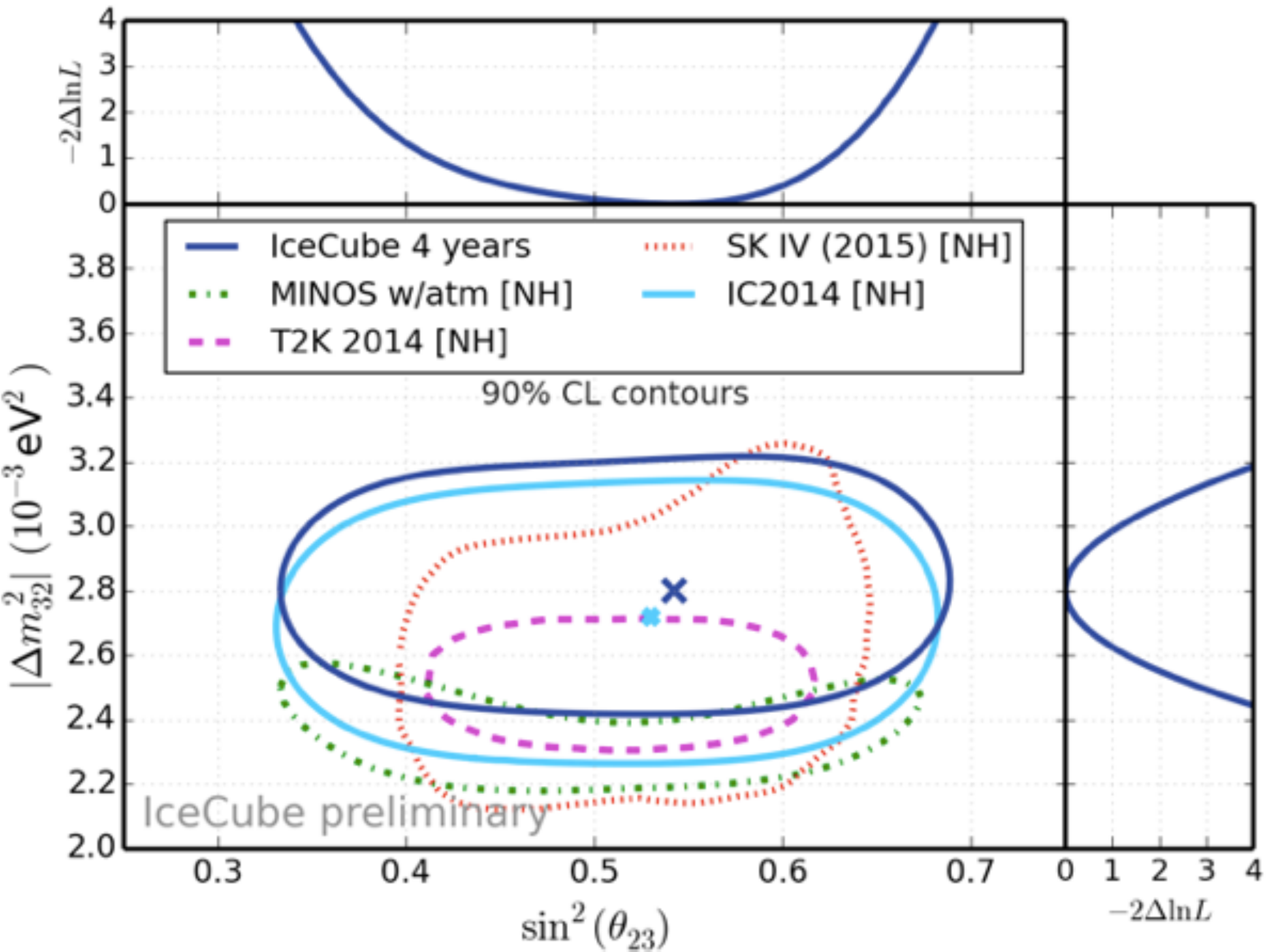}
    \caption{\label{fig:osci_icecube} The preferred 90\% CL region in the oscillation parameter space of atmospheric neutrinos measured by DeepCore in 4 yrs
with the region published in \cite{osci-dc} with a sample of 3 yrs. The best fit parameters are $\Delta m^2 = 2.80^{+0.20}_{-0.16} \times 10^{-3}$ eV$^2$ and $\sin^2 2\theta = 0.54^{+0.08}_{-0.13}$.} 
\end{figure}

Neutrino oscillations are, together with dark matter, another topic where the action is in the relatively low energy range of IceCube and 
for which in-fill arrays like DeepCore and the proposed PINGU \cite{PINGU} are extremely relevant to increase detection power.
The three-year measurement of standard oscillation parameters published in Ref.~\cite{osci-dc} has been recently updated with the addition of one more year.
About 6400 events with energies between 6-56 GeV measured by DeepCore are used to obtain the result shown in Fig.~\protect\ref{fig:osci_icecube}.
The first maximum of $\nu_{\mu}$ disappearance happens at 25 GeV of neutrino energy.
\begin{figure}[htb]
    \includegraphics[width=0.47\textwidth]{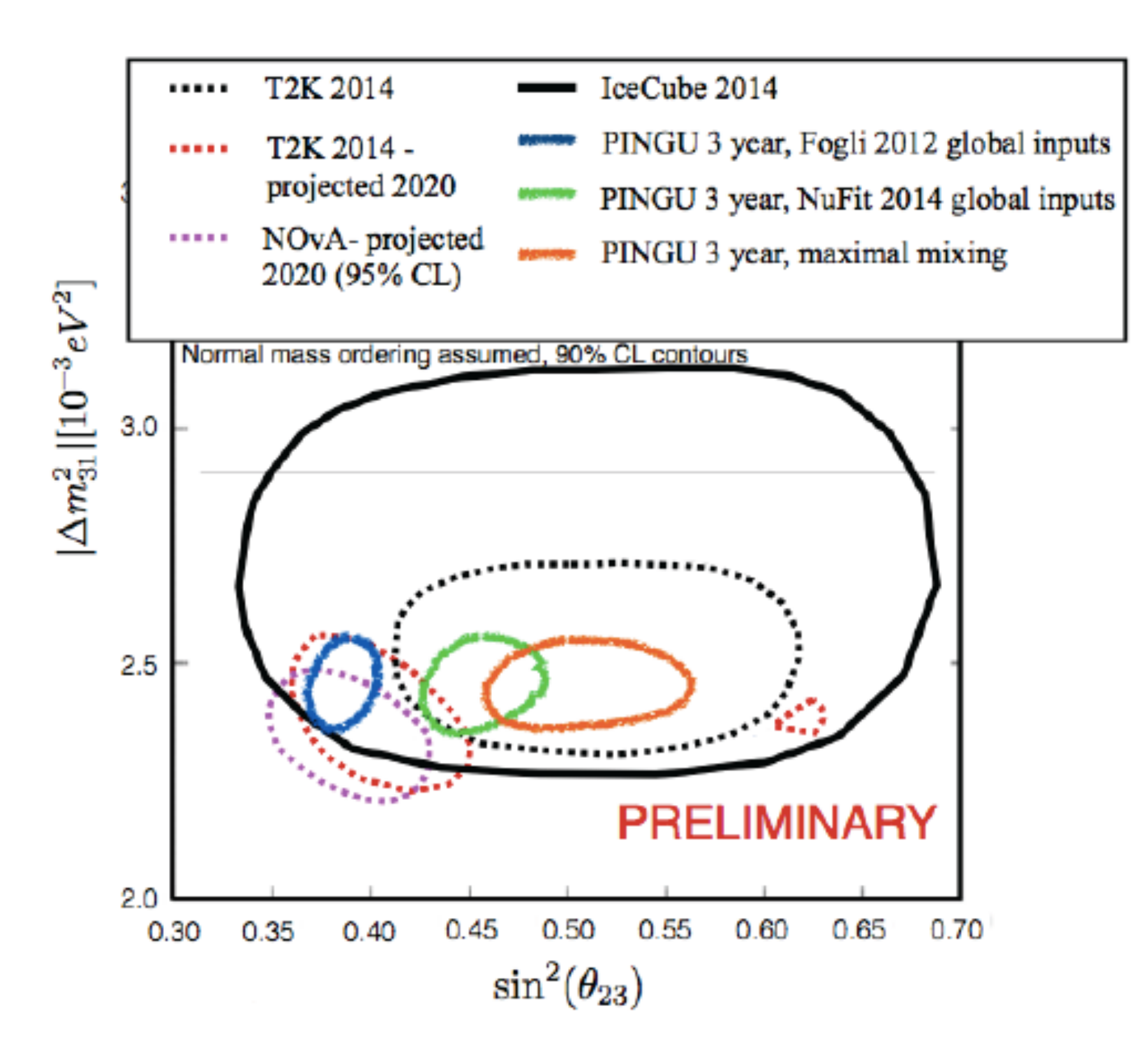}
    \caption{\label{fig:PINGU} The IceCube sensitivity in the oscillation parameter space \protect\cite{osci-dc}, the T2K achieved, the T2K and NOVA projected sensitivities are shown 
with the PINGU projected sensitivity~\protect\cite{gen2} at comparable time.} 
\end{figure} 

\begin{figure}[htb]
    \includegraphics[width=0.47\textwidth]{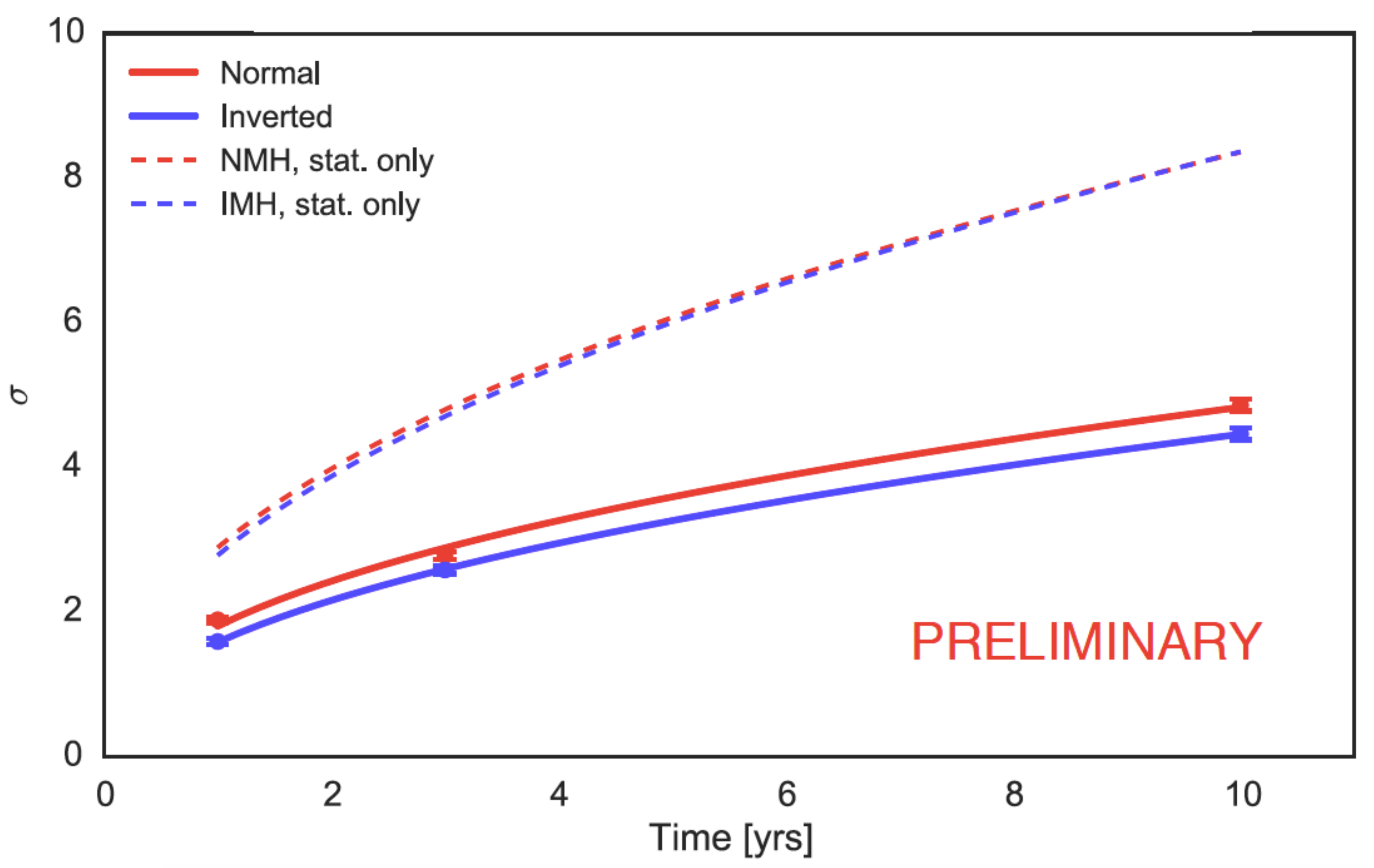}
    \caption{\label{fig:significance} The evolution of the significance for determining the neutrino mass hierarchy with the PINGU
detector vs years~protect\cite{gen2}.} 
\end{figure}

The sensitivity of PINGU is compared to current results in Fig.~\protect\ref{fig:PINGU} and the potential to discriminate the hierarchy of neutrino masses is shown in  Fig.~\protect\ref{fig:significance} \cite{gen2}.
IceCube and the future PINGU have sensitivity in the region of sterile neutrinos with masses in the eV range which could explain the LSND anomaly. Hence, they are complementary to
the US short baseline program.

\section{The future of IceCube: IceCube-Gen2}

Recently, the IceCube Collaboration largely expanded to include groups interested in working in a bi-directional expansion of IceCube, IceCube-Gen2 \cite{gen2}. The low energy expansion 
is PINGU \cite{PINGU} and its goals have been described in the two previous sections.
A high-energy expansion requires on one side an expansion of the deep detector and on the other a large surface veto to improve the detection of starting events. The increased statistics will improve
the understanding of their spectrum and will allow to identify their origin. A sufficiently large array would also allow the capability to detect neutrinos
produced in the ice layer between the deep detector and the surface array.   
A number of possible solutions for cheap units at the surface have been considered, such as water tanks or scintillators equipped with photosensors or
1m-dish air Cherenkov telescopes.




\nocite{*}
\section{Acknowledgements}
The research on which this paper reports
is supported in part by the U.S. National Science
Foundation and by the Swiss SNF. A full list of supporting agencies for
the IceCube Neutrino Observatory may be found at http://icecube.wisc.edu/collaboration/funding.
\bibliographystyle{elsarticle-num}



\end{document}